# Error Performance Analysis to Increase Capacity of A Cellular System Using SDMA

Md. M. Hossain, J. Hossain

**Abstract**— One of the biggest drawbacks of the wireless environment is the limited bandwidth. However, the users sharing this limited bandwidth have been increasing considerably. Space Division Multiple Access (SDMA) is a new technology by which the capacity of existing mobile communication systems can economically be increased. This paper has been presented how the capacity can be enhanced by using SDMA with smart antennas in mobile communications system. Based on Adaptive Antenna Array (AAA) technology the spatial dimension of the existing system is exploited by means of forming independent radio beams in each of the original channels. This paper analyses the comparison of average Bit Error Rate (BER) of SDMA and CDMA technique and the different ways in which SDMA can be introduced to increase the capacity of a cellular system. The probability of error is found for a standard omni directional base station antenna, and another set of curves is found for flat top beam having a directivity of 5.1dB. It is assumed that k separate flat top beams can be formed by base station and pointed each of the k users within the cell of interest. Noticing that for an average probability of error greater than 0.1 in a propagation path loss environment of n=4, the flat top beam will support 200 users, whereas the omni-directional antenna will support only 50 users. This increase the number of user is roughly equal to the directivity offered by the flat top beam system, and illustrates the promise SDMA offers for improving capacity in wireless system. Here multipath fading is not considered.

———————————— ◆ ————————————

## 1 INTRODUCTION

Recent advances in the area of wireless communications have revealed the emerging need for efficient wireless access in personal, local and wide area networks. SDMA with smart antennas at the base station is recognized as a promising means of increasing system capacity and supporting rate-demanding services. Mobile radio communication systems are currently characterized by an ever-growing number of users, which however is coupled with limited available resources, in particular in terms of usable frequency spectrum. The SDMA technique allows enhancing the capacity of a cellular system by exploiting spatial separation between users. In an SDMA system, the base station does not transmit the signal throughout the area of the cell, as in the case of conventional access techniques, but rather concentrates power in the direction of the mobile unit the signal is meant to reach, and reduces power in the directions where other units are present. The same principle is applied for reception. The SDMA technique has several characteristics which make its introduction in a mobile radio system advantageous. In particular, all modifications required are limited to base stations, and do not involve mobile units. Moreover, the SDMA technique can be integrated with different multiple access techniques (FDMA, TDMA, CDMA), and therefore it can be used in all mobile radio systems currently operating or about to be introduced.

In the last couple of years there has been large activity in the field of SDMA system just applying separate antenna in a Base Station (BS) for sectoring a cell. In most of those works there have been shown the relation between average BER and Signal to Noise Ratio (SNR), Cumulative Distribution Function (CDF) of Carrier to Interference Ratio (C/I) for a Spatial Filtering for Interference Reduction (SFIR), average BER with C/I in SDMA etc. But in this thesis the work is done on the average BER analysis with the users in SDMA using smart antenna for mobile communications.

## 2 OVERVIEW OF SDMA

In SDMA a number of users share the same available resources and are distinguished only in the spatial dimension. In traditional cellular systems the base station, having no information on the position of mobile units, is forced to radiate the signal in all direction to cover the entire area of the cell. This entails both a waste of power and the transmission, in the directions where there are no mobile terminals to reach, of a signal which will be seen as interfering for co-channel cells, i.e. those cells using the same group of radio channels. Analogously, in reception, the antenna picks up signals coming from all directions, including noise and interference.

These considerations have lead to the development of the SDMA technique, which is based on deriving and exploiting information on the spatial position of mobile terminals. In particular, the radiation pattern of the base station, both in transmission and reception is adapted to each different user so as to obtain, as shown in Fig. 1, the highest gain in the direction of the mobile user. Simultaneously, radiation nulls shall be positioned in the directions of interfering mobile units. This behavior is just defined "Interference nulling" [3], [5].

————————————————

• *Md. M. Hossain is with the Electronics and Communication Engineering Discipline, Khulna University, Khulna -9208, Bangladesh.*
*J.Hossain. is with the Axiata Limited Bangladesh(GSM operator).*

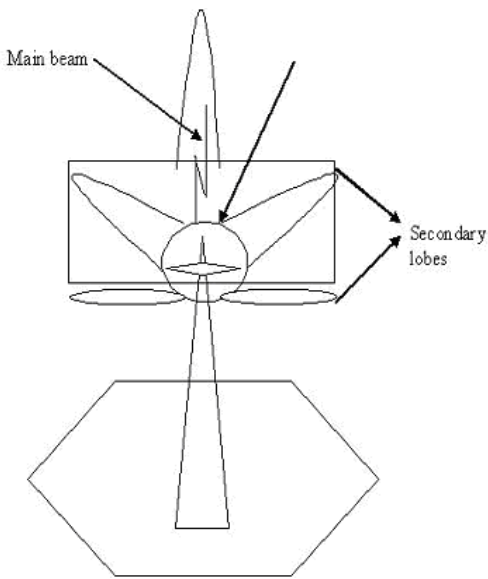

Fig. 1. Radiation beam pattern adapted for a single user

the highest gain in the direction of the mobile user. Simultaneously, radiation nulls shall be positioned in the directions of interfering mobile units. This behavior is just defined "Interference nulling" [3],[5].

SDMA is widely recognized as one of the most promising techniques for improving the capacity of cellular systems. This access technique allows different users to be served on the same frequency channel at the same time thus improving the spectral efficiency. SDMA can be implemented by the use of antenna arrays at the base station, and by exploiting the possibility to manage intracell interference with appropriate signal processing techniques like beamforming or linear preceding [8]. On the other hand, and differently from TDMA or FDMA, SDMA is intrinsically affected by interference among users that share the same resources. A fundamental issue when designing SDMA schemes consists in the possibility to control and foresee the degree of interference experienced by spatially multiplexed users.

In a SDMA mobile radio system, different users are separated by Directions of Arrival [DOAs] of their signals at the base station [9]. Thus, a part of the spectrum can be used not only by one user at a certain time instant, as in conventional TDMA or FDMA systems, but by several users. The condition is that the different signals arriving at the base station are spatially well separated. The main additional task compared to conventional systems that has to be performed in a SDMA system is the estimation of the DOAs. When the users move around, the DOAs are time variable, which means that they must be tracked after their initial estimation.

## 3 SMART ANTENNAS

A smart antenna is an antenna system that is able to direct the beam at each individual user, allowing the users to be separated in the spatial domain. It is not the antenna that is smart, but the antenna system. The impact of using smart antennas depend both on the level of intelligence of the antenna system and the type of mobile system in which it is deployed [5].

The utilization of smart antennas in mobile systems can be divided in 2 main stages.
These are:
Spatial Filtering for Interference Reduction
Adaptive Antenna Array

### 3.1 SPATIAL FILTERING FOR INTERFERENCE REDUCTION (SFIR)

SFIR uses the beam directivity from smart antennas to reduce the interference [3],[10]. In GSM, this is a TDMA/FDMA system, this interference reduction results in an increase of the capacity or the quality in the system. This is achieved by either allowing a tighter re-use factor and thereby a higher capacity, or to keep the same re-use factor but with a higher SNR level and signal quality (i.e. if the current re-use factor is 7/21, the same SNR could be achieved with SFIR and a re-use factor of 3/9) [4].

### 3.2 ADAPTIVE ANTENNA ARRAY (AAA)

In adaptive antenna array system the beamforming is done digitally, and a main lobe is generated in the direction of the strongest signal component [6],[7]. In addition, side lobes are generated in the direction of multi path components and nulls in the direction of interferers. This technique will maximize the SNR. The operation of an adaptive antenna array [5] at the base station is illustrated in Fig 2.

As in the Fig. 2 this system comprises an array of antennas and a Digital Signal Processor (DSP) tasked with real time processing of signals received or to be sent to the different antennas. With reference to the Fig. 2, it can be observed that the RF signal received by each of the N antennas comprising the array is at first brought down to base band and then converted into digital form. The N complex signals obtained are then sent as inputs to the DSP, which multiplies the signal of each antenna by a suitable factor wi*, and finally adds the various terms. The output signal is therefore given by [5]:

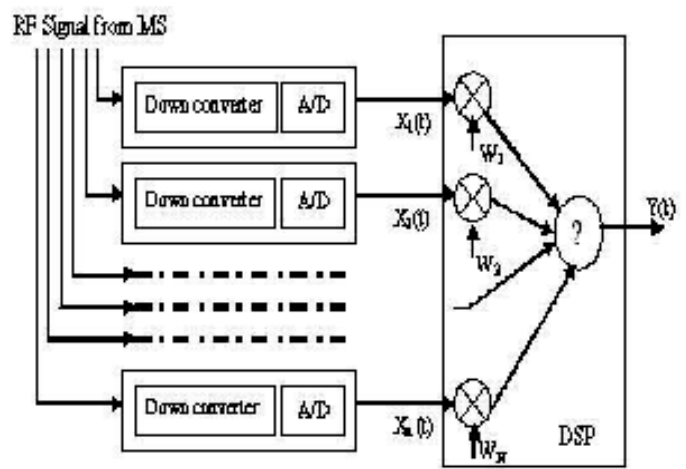

Fig. 2. Structure of an adaptive antenna array in reception

$$y(t) = \sum_{i=1}^{N} W_i * X_i(t) \qquad (1)$$

An appropriate choice of the weights vector w = [$w_1$, $w_2$, ..., $w_N$] allows to give the radiation pattern the desired characteristics. In particular, vector w is determined using an adaptive strategy. Coefficients are therefore updated periodically, based on the observation of data received. To assure correct operation of the system, it is necessary that the adoption rate could compensate the environmental variations, due to the mobility of the sources and accentuated by the presence of multiple paths. The use of an adaptive antenna array at the base station thus allows introducing the SDMA technique, whose main advantage is the capability to increase system capacity, i.e. the number of users it can handle.

## 4 CHANNEL CAPACITY

### 4.1 CAPACITY OF CODE DIVISION MULTIPLE ACCESS (CDMA)

For interference limited CDMA operating is an Additive White Gaussian Noise(AWGN) channel, with perfect power control with no interference from adjacent cells & with omni directional antennas used at the base station, the average bit error rate, $P_b$, for a user can be found from the Gaussian approximation as [6],[1]:

$$P_b = Q\left(\sqrt{\frac{3N}{k-1}}\right) \quad (2)$$

where k is the number of user in a cell and N is the spreading factor. Q(x) is the standard Q-function.

### 4.2 SIMULATION OF AVERAGE BER FOR CDMA (USING MATLAB)

For CDMA operating is an AWGN channel, with perfect power control with no interference from adjacent cells & with omni directional antennas used at the base station, the average bit error rate, $P_b$, for a user can be found from the equation 2

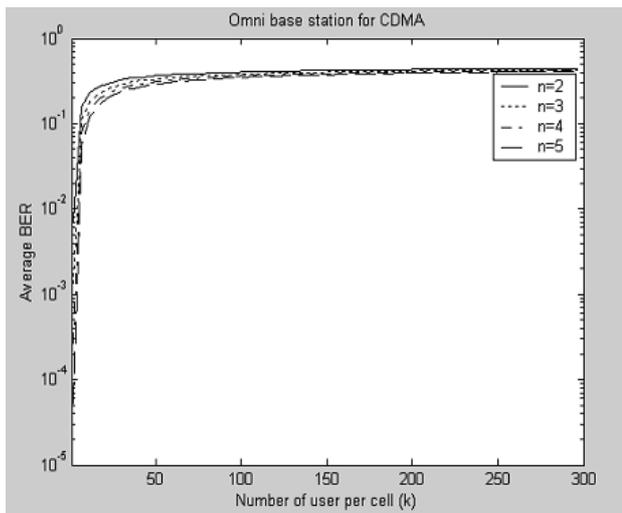

Fig. 3. Average probability of error for CDMA system with one interfering co-channel cells.

### 4.3 CAPACITY OF SPACE DIVISION MULTIPLE ACCESS (SDMA)

The average bit error rate for SDMA system is given by [1]:

$$P_b = Q\sqrt{\left(\frac{3DN}{k-1}\right)} \quad (3)$$

where D is the directivity of the antenna.

### 4.4 SIMULATION OF AVERAGE BER FOR SDMA (USING MATLAB)

The average bit error rate, $P_b$, for a user can be found from the equation 3

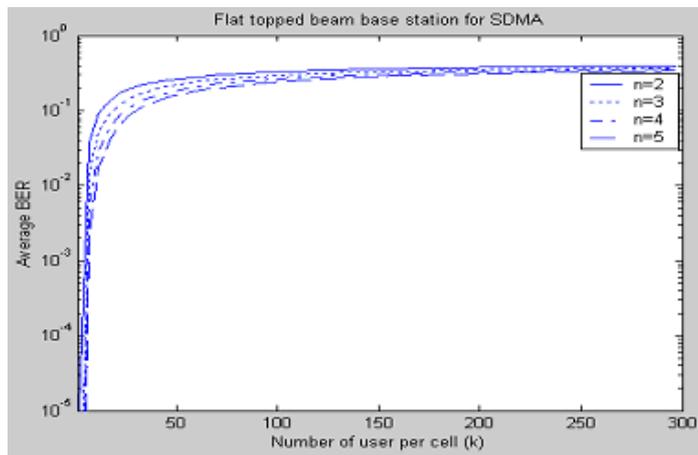

Fig . 4. Average BER for SDMA system with one interfering co-channel cells

### 4.5 SIMULATION OF AVERAGE BER FOR THE COMPARISON BETWEEN CDMA AND SDMA (USING MATLAB)

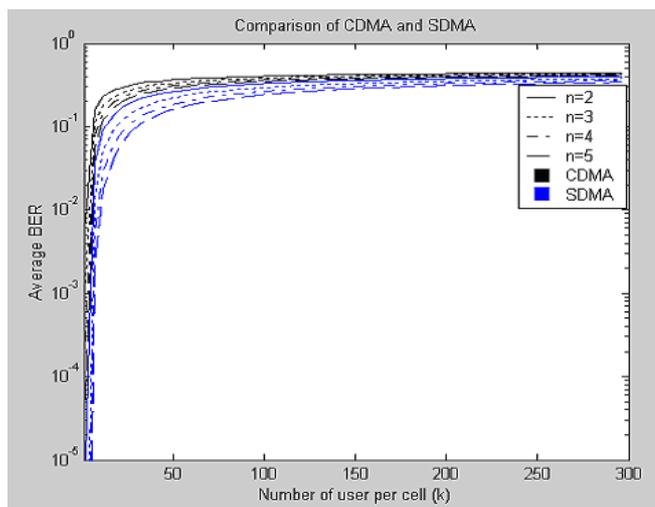

Fig. 5. Comparison of average probability of error between CDMA and SDMA

Above Fig 5, illustrates the average probability of error for dif-ferent propagation path loss exponent, where two different types of base antennas are compared using simulation.

## 4.6 CAPACITY INCREASE IN SDMA TECHNIQUE

By using SDMA technique capacity increase can be obtained in two different ways, and therefore the following applications are possible:
Reduction in Co-channel Interference
Spatial Orthogonality

### 4.6.1 REDUCTION IN CO-CHANNEL INTERFERENCE

The reduction in the level of co-channel interference between the different cells using the same group of radio channels is obtained by minimizing the gain in the direction of interfering mobile units. This technique indicated with the acronym SFIR allows reducing frequency re-use distance and clustering size. In this way, each cell can be assigned a higher number of channels [2],[3].

### 4.6.2 SPATIAL ORTHOGONALITY

In conventional access techniques, Orthogonality between signals associated with different users is obtained by transmitting them in different frequency bands (FDMA), in different time slots (TDMA) or using different code sequences (CDMA). Using an adaptive antenna array, it is possible to create an additional degree of Orthogonality between signals transmitted to and from different directions. It is thus possible to assign the same physical channel to several mobile units, when the angles at which they are seen by the base station are sufficiently separated. The result is an increase in the number of available channels, since the same physical channel, for example the same carrier in a FDMA system or the same time slot in a TDMA system, can be subdivided into multiple spatial channels, each of which is assigned to a different user.

## 5 RESULT

From Fig 3, it was found that the average probability of error for a user in a CDMA system with one interfering co-channel cells using omni directional base station antenna, which has directivity, D=0 dB. Also from Fig 4, it was found that the average probability of error for a user in a SDMA system with a flat top beam, which has directivity, D=5.1 dB.

In both Figure 4 and Figure 5, n is path loss exponent and k was taken as user within the cell of interest, assuming that k=300.

In Fig 5, SDMA offered significant capacity gains for a given average BER in comparison with CDMA.

## 6 CONCLUSION

In Fig 5 one set of probability of error is found for a standard omni directional base station antenna, and another set of curves is found for flat top beam (a beam with constant gain over a specific region) having a directivity of 5.1dB. It is assumed that k separate flat top beams can be formed by base station and pointed each of the k users within the cell of interest. Noticing that for an average probability of error greater than 0.1 in a propagation path loss environment of n=4, the flat top beam will support 200 users, whereas the omni-directional antenna will support only 50 users.

This increase the number of user is roughly equal to the directivity offered by the flat top beam system, and illustrates the promise SDMA offers for improving capacity in wireless system. Here multipath fading is not considered.

The impact of scattering and diffuse multipath on the performance of SDMA is currently a topic of research and is certain to impact performance and implementation strategies for emerging SDMA techniques.

**REFERENCES**

[1] Theodore S. Rappaport (2007),*"Wireless Communications Principles And Practice"*- pp. 621-643,Second Edition,Chapter.9,Person Education (Sinagapore).

[2] Amir Leshem and Sharon Gannot, "Robust Sequential Interference Cancellation For Space Division Multiple Access Communication", School of Electrical Engineering Bar-Ilan University, Ramat-Gan 52900, Israel.

[3] Arild Jacobsen, *"SMART ANTENNAS for DUMMIES"*, © Telenor AS 25/07/00.

[4] B. D. Van Veen, K. M. Buckley, "Beamforming: A Versatile Approach to Spatial

Filtering," *IEEE ASSP Magazine, April 1988, pp.4-24.*

[5] Enrico Buracchini, "SDMA IN MOBILE RADIO SYSTEMS: CAPACITY ENHANCEMENT IN GSM & I 95".

[6] LAL C. GODARA, SENIOR MEMBER, IEEE, "Applications of Antenna Arrays to Mobile Communications, Part I: Performance Improvement, Feasibility, and System Considerations".

[7] M. Tangemann et alii, "Introducing Adaptive Array Antenna Concepts in Mobile Communication Systems," in Proceedings RACE Mobile Communications Workshop, Amsterdam, The Netherlands, May 1994, pp. 714-727.

[8] Notker Gerlich and Michael Tangemannz, "Towards a Channel Allocation Scheme for SDMA_based Mobile Communication Systems" University of Wurzburg Institute of Computer Science Research Report Series, Report No-104, February 1995.

[9] Rupert Stutzle, Jurgen Gotze, and Josef A. Nossek , "Updating Directions of Arrival in a GSM-Based SDMA MobileRadio System" Institute for Network Theory and Circuit Design, Munich University of Technology Arcisstr. 21, D-80290 Muenchen, Germany.

[10] Alexander Kuchar, Josef Fuhl, Ernst Bonek, "Spectral Efficiency Enhancement and Power Control of Smart Antenna Systems", Institut fur Nachrichtentechnik und Hochfrequenztechnik, Technische University at Wien, Vienna, Gusshausstrasse 25/389, A-1040 Wien, Austria.

**Md. M. Hossain** received his B.Sc Engineering degree in Electronics and Communication in the year of 2003 from Khulna University,Khulna-9208,Bangladesh.He is now the faculty member of Electronics and Communication Enfgineering Discipline,Khulna University,Khulna-9208,Bangladesh.His current research interest is wireless communication,modulation,channel coding and fading. His number of published paper are 3 among them one was published in the proceding of 12[th] IC-CIT,2009,Dhaka,Bangladesh.Another one was published in the proceding of 2[nd] international conference on ICEESD,2009,DHAKA,Bangladesh and last one was published in the conference on Engineering Research,Innovation and Education,2010,SUST,Sylhet,Bangladesh.

**J. Hossain** received his B.Sc Engineering degree in Electronics and Communication in the year of 2007 from Khulna University,Khulna-9208,Bangladesh.He is now working in Axiata Bangladesh(GSM operator) Lld.as Engineer,NSS core network operation.